\documentclass[runningheads]{llncs}

\usepackage[T1]{fontenc}
\def\doi#1{\href{https://doi.org/\detokenize{#1}}{\url{https://doi.org/\detokenize{#1}}}}
\usepackage{graphicx}
\usepackage{listings}
\usepackage{graphicx}
\graphicspath{{figures}}
\usepackage{makecell}
\usepackage{bm}
\usepackage{multirow}
\usepackage{footmisc}
\usepackage[table]{xcolor}
\usepackage{hhline}
\usepackage{wrapfig}
\usepackage{array}

\newcounter{mysfig}
\counterwithin{mysfig}{figure}

\renewcommand\themysfig{(\alph{mysfig})}
\makeatletter
\newcommand\Scaption[1]{%
\refstepcounter{mysfig}%
\vskip.5\abovecaptionskip
  \sbox\@tempboxa{\small\themysfig~#1}%
  \ifdim \wd\@tempboxa >\hsize
    \small\themysfig~#1\par
  \else
    \global \@minipagefalse
    \hb@xt@\hsize{\hfil\box\@tempboxa\hfil}%
  \fi
  \vskip\belowcaptionskip}
\makeatother

\newlength{\Oldarrayrulewidth}

\newcolumntype{C}[1]{>{\centering}m{#1}}
\begin{document}

\title{Mixing Backward- with Forward-Chaining for Metacognitive Skill Acquisition and Transfer}

\titlerunning{Mixing Backward- with Forward-Chaining}

\author{Mark Abdelshiheed \and John Wesley Hostetter \and Xi Yang \and \\Tiffany Barnes \and Min Chi\\ \{mnabdels,\, jwhostet,\, yxi2,\, tmbarnes,\, mchi\}@ncsu.edu}

 \authorrunning{Abdelshiheed et al.}

\institute{North Carolina State University, Raleigh, NC 27695, USA}

\maketitle              % typeset the header of the contribution
\begin{abstract}
Metacognitive skills have been commonly associated with preparation for future learning in deductive domains. Many researchers have regarded \emph{strategy-} and \emph{time-awareness} as two metacognitive skills that address \emph{how} and \emph{when} to use a problem-solving strategy, respectively. It was shown that students who are both strategy- and time-aware $(StrTime)$ outperformed their $nonStrTime$ peers across deductive domains. In this work, students were trained on a logic tutor that supports a default forward-chaining (FC) and a backward-chaining (BC) strategy. We investigated the impact of mixing BC with FC on teaching strategy- and time-awareness for $nonStrTime$ students. During the logic instruction, the experimental students $(Exp)$ were provided with two BC worked examples and some problems in BC to practice \emph{how} and \emph{when} to use BC. Meanwhile, their control $(Ctrl)$ and $StrTime$ peers received no such intervention. Six weeks later, all students went through a probability tutor that only supports BC to evaluate whether the acquired metacognitive skills are transferred from logic. Our results show that on both tutors, $Exp$ outperformed $Ctrl$ and caught up with $StrTime$.

\end{abstract}

\section{Introduction}

One fundamental goal of education is being prepared for future learning \cite{bransford1999transferRethinking} by transferring acquired skills and problem-solving strategies across different domains. Despite the difficulty of achieving such transfer \cite{bransford1999transferRethinking}, prior research has shown it can be facilitated by obtaining metacognitive skills \cite{abdelshiheed2022power,abdelshiheed2022assessing,abdelshiheed2021preparing,abdelshiheed2020metacognition,chi2010backward2metacogStrategyINSTRUCTION}. It has been believed that metacognitive skills are essential for academic achievements \cite{de2018longSWITCH-INSTRUCTION}, and teaching such skills impacts learning outcomes \cite{chi2010backward2metacogStrategyINSTRUCTION} and strategy use \cite{schraw2015metacognitiveInstruction}. Much prior research has categorized knowing \emph{how} and \emph{when} to use a problem-solving strategy as two metacognitive skills \cite{winne2014switchMetacognitiveSWITCH}, referred to as strategy- and time-awareness, respectively. Our prior work found that students who were both strategy- and time-aware ---referred to as $StrTime$--- outperformed their $nonStrTime$ peers across deductive domains \cite{abdelshiheed2021preparing,abdelshiheed2020metacognition}. In the current work, we provide interventions for the latter students to catch up with their $StrTime$ peers.

\noindent Deductive domains such as logic, physics and probability usually require multiple problem-solving strategies. Two common strategies in these domains are forward-chaining (FC) and backward-chaining (BC). Early studies showed that experts often use a mixture of FC and BC to execute their strategies \cite{priest1992mixFCBC}. This work investigates the impact of mixing FC and BC on teaching strategy- and time-awareness for $nonStrTime$ students.

Our study involved two intelligent tutoring systems (ITSs): logic and probability. Students were first assigned to a logic tutor that supports FC and BC, with FC being the default, then to a probability tutor six weeks later that only supports BC. During the logic instruction, $nonStrTime$ students were split into experimental $(Exp)$ and control $(Ctrl)$ conditions. For $Exp$, the tutor provided two worked examples solved in BC and presented some problems in BC to practice \emph{how} and \emph{when} to use BC. $Ctrl$ received no such intervention as each problem was presented in FC by default with the ability to switch to BC. Our goal is to inspect whether our intervention would make $Exp$ catch up with the golden standard ---$StrTime$ students--- who already have the two metacognitive skills and thus need no intervention. All students went through the same probability tutor to evaluate whether the acquired metacognitive skills are transferred from logic. Our results show that $Exp$ outperformed $Ctrl$ and caught up with $StrTime$ on both tutors.

\subsection{Metacognitive Skill Instruction}

Metacognitive skills regulate one's awareness and control of their  cognition \cite{chambres2002metacognitionStrategySelection}. Many studies have demonstrated the significance of metacognitive skills instruction on academic performance \cite{de2018longSWITCH-INSTRUCTION}, learning outcomes \cite{abdelshiheed2022power,abdelshiheed2021preparing,chi2010backward2metacogStrategyINSTRUCTION} and regulating strategy use \cite{schraw2015metacognitiveInstruction}. Schraw and Gutierrez \cite{schraw2015metacognitiveInstruction} argued that metacognitive skill instruction involves feeling what is known and not known about a task. They stated that such instruction should further compare strategies according to their feasibility and familiarity from the learner's perspective. Chi and VanLehn \cite{chi2010backward2metacogStrategyINSTRUCTION} found that teaching students principle-emphasis skills closed the gap between high and low learners, not only in the domain where they were taught (probability) but also in a second domain where they were not taught (physics).

Strategy- and time-awareness have been considered metacognitive skills as they respectively address \emph{how} and \emph{when} to use a problem-solving strategy \cite{de2018longSWITCH-INSTRUCTION,winne2014switchMetacognitiveSWITCH}. Researchers have emphasized the role of strategy awareness in preparation for future learning \cite{abdelshiheed2021preparing,belenky2012strategyAwarenessPFL} and the impact of time awareness on planning skills and academic performance \cite{de2018longSWITCH-INSTRUCTION,fazio2016timeAwareness}. Belenky and Nokes \cite{belenky2012strategyAwarenessPFL} showed that students who had a higher aim to master presented materials and strategies outperformed their peers on a transfer task. Fazio et al. \cite{fazio2016timeAwareness} revealed that students who knew when to use each strategy to pick the largest fraction magnitude had higher mathematical proficiency than their peers. de Boer et al. \cite{de2018longSWITCH-INSTRUCTION} showed that students who knew when and why to use a given strategy exhibit long-term metacognitive knowledge that improves their academic performance. de Boer et al. emphasized that knowing \emph{when} and \emph{why} has the same importance as knowing \emph{how} when it comes to strategy choice in multi-strategy domains.

\subsection{Forward- and Backward-Chaining}

FC and BC are two standard problem-solving strategies in deductive domains. In FC, the reasoning proceeds from the given propositions toward the target goal, whereas BC is goal-driven in that it works backward from a goal state to a given state. Substantial work has investigated the impact of FC and BC strategies in two research categories: empirical studies and post-hoc observations.

Prior empirical studies have shown the significance of FC over BC in learning physics \cite{larkin1980expertFC-BC} and weightlifting movements \cite{moore2019FC-BC}. Moore and Quintero \cite{moore2019FC-BC} compared FC and BC in teaching the clean and snatch movements to novice weight lifters. The participants showed mastery performance with the FC training but showed substantially fewer improvements in performance accuracy via the BC training. All participants mastered the movements when some BC lifts were changed to FC. Conversely, some studies reported no significant difference between the two strategies \cite{slocum2011assessmentFC-BC}. Slocum and Tiger \cite{slocum2011assessmentFC-BC} assessed the children's FC and BC strategy preferences on various learning tasks. They found that children were equally efficient on both strategies and had similar mixed strategy preferences. 

Early research has observed the impact of mixing FC and BC strategies \cite{priest1992mixFCBC}. Priest and Lindsay \cite{priest1992mixFCBC} compared how experts and novices solve physics problems. Although both groups used a mixture of FC and BC, \emph{only} the experts knew how and when to use each strategy and significantly produced more complete plans and stages than their novice peers. In brief, while no consensus has been reached on whether FC or BC is most effective in problem-solving, prior work has observed that the mixture of FC and BC yields the highest performance accuracy as learners know how and when to use each strategy.

\section{Methods}

\noindent{\textbf{Participants}} They are Computer Science undergraduates at North Carolina State University. Students were assigned each tutor as a class assignment and told that completion is required for full credit. Similar to our prior work, we utilize the random forest classifier (RFC) that, based on pre-test performance, predicts the metacognitive label ($StrTime$ or otherwise) before training on logic and was previously shown to be $96\%$ accurate \cite{abdelshiheed2021preparing}. Specifically, $StrTime$ students frequently follow the desired behavior of switching \emph{early} (within the first $30$ actions) to $BC$, while their peers either frequently switch late (after the first $30$ actions) or stick to the default $FC$ \cite{abdelshiheed2022power,abdelshiheed2021preparing,abdelshiheed2020metacognition}. A total of $121$ students finished both tutors and were classified by the RFC into $26$ $StrTime$ and $95$ otherwise. The latter students were randomly assigned to $Experimental$ $(Exp: N = 49)$ and $Control$ $(Ctrl: N=46)$ conditions. The RFC was $97\%$ accurate in classifying students who received no intervention ---$Ctrl$ and $StrTime$.

\begin{figure}[ht!]
\begin{minipage}[t]{0.28\columnwidth}
\begin{minipage}[b]{1\columnwidth}
\includegraphics[width=1\linewidth, height=4.1cm]{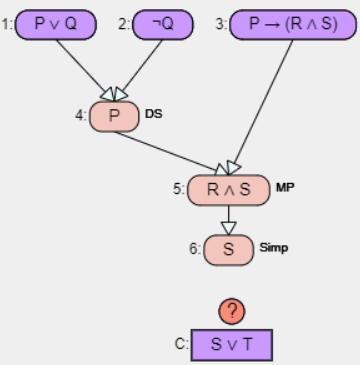}\par
\end{minipage}
\Scaption{FC Strategy}
\label{fig:fc}
\end{minipage}\hfill
\begin{minipage}[t]{0.36\columnwidth}
\begin{minipage}[b]{\columnwidth}
\includegraphics[width=\linewidth, height=4.1cm]{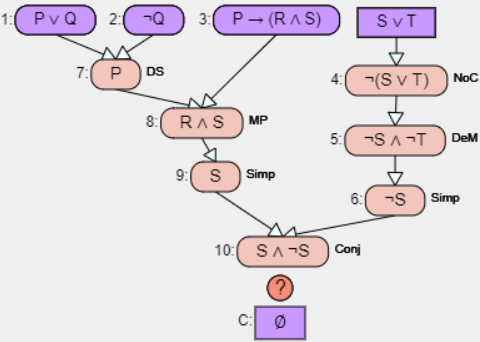}
\end{minipage}
\Scaption{BC Strategy}
\label{fig:bc}
\end{minipage}\hfill
\begin{minipage}[t]{0.32\columnwidth}
\begin{minipage}[b]{1\columnwidth}
\includegraphics[width=1\linewidth,height=4.1cm]{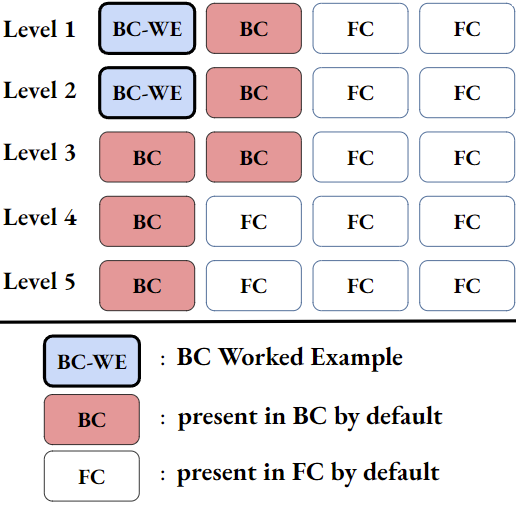}\par
\end{minipage}
\Scaption{Modified Training}
\label{fig:modifiedTraining}
\end{minipage}
\vskip0.1in
\caption{Logic Tutor} 
\label{fig:logic}
\end{figure}

\vskip 0.1 in
\noindent{\textbf{Logic Tutor and Our Intervention}}
The logic tutor teaches propositional logic proofs by applying inference rules such as Modus Ponens. A student can solve any problem by either a \textbf{FC} or \textbf{BC} strategy. Students derive a conclusion at the bottom from givens at the top in \emph{FC} (Fig. 1a), while they derive a contradiction from givens and the \emph{negation} of the conclusion in BC (Fig. 1b). A problem is presented by \emph{default} in FC with the ability to switch to BC by clicking a button. The tutor consists of two pre-test, $20$ training and six post-test problems. The post-test is \emph{much harder} than the pre-test, and the first two post-test problems are isomorphic to the two pre-test problems. The \emph{pre-} and \emph{post-test} scores are calculated by averaging the pre- and post-test problem scores, where a problem score is a function of time, accuracy, and solution length. The training consists of five ordered levels in an \emph{incremental degree of difficulty}, and each level consists of four problems. We modified the training section to mix BC with FC (Fig. 1c). Specifically, two worked examples (WE) on BC were implemented, where the tutor provided a step-by-step solution, and six problems were presented in BC by default. The two WEs and the six problems are expected to teach students \emph{how} and \emph{when} to use BC. Note that the colored problems in Figure 1c were selected based on the historical strategy switches in our data \cite{abdelshiheed2020metacognition}.

\vskip 0.07 in
\noindent{\textbf{Probability Tutor}} It teaches how to solve probability problems using ten principles, such as the Complement Theorem. The tutor consists of a textbook, pre-test, training, and post-test. The textbook introduces the domain principles, while training consists of $12$ problems, each of which can \emph{only} be solved by $BC$ as it requires deriving an answer by \emph{writing and solving equations} until the target is ultimately reduced to the givens. In pre- and post-test, students solve $14$ and $20$ open-ended problems graded by experienced graders in a double-blind manner using a partial-credit rubric. The \emph{pre-} and \emph{post-test} scores are the average grades in their respective sections, where grades are based \emph{only} on accuracy. Like the logic tutor, the post-test is much harder than the pre-test, and each pre-test problem has a corresponding isomorphic post-test problem.

\vskip 0.1 in
\noindent{\textbf{Procedure}} Students were assigned to the logic tutor and went through the pre-test, training and post-test. Before training on logic, the RFC predicted the metacognitive label for each student, as described in the Participants section. During training, $Exp$ received the modified tutor shown in Figure 1c, while $Ctrl$ and $StrTime$ received the original tutor, where all problems are presented in $FC$ by default. Six weeks later, students were trained on the probability tutor.

\section{Results}

\begingroup
\begin{table}[ht!]
\scriptsize
\begin{center} 
\caption{Comparing Groups across Tutors} 
\label{groupSummary} 
\begin{tabular}{cC{2.75cm}C{2.75cm}|c} 
\Xhline{4\arrayrulewidth}

    & \makecell[t]{$Experimental\, (Exp) $ \\ $(N=49)$}
    & \makecell[t]{$Control \, (Ctrl)$ \\ $(N=46)$} & \makecell[t]{$StrTime$ \\ $(N=26)$} \\

\hline
\multicolumn{4}{c}{Logic Tutor}\\
\hline
$Pre$ &  $61.7\,(18)$ &  $58.7\,(20)$ & \cellcolor{gray!40}$62.1\,(20)$ \\
$Iso$-$Post$ &  $81\,(11)$ &  $70.4\,(14)$ & \cellcolor{gray!40} $81.3\,(10)$ \\
$Iso$-$NLG$ &  $0.27\,(.12)$ &  $0.09\,(.31)$ & \cellcolor{gray!40} $0.29\,(.16)$ \\
$Post$ &  $77.4\,(11)$  &   $66.7\,(14)$ & \cellcolor{gray!40} $79\,(9)$ \\
$NLG$ &  $0.24\,(.15)$ &   $0.06\,(.37)$ & \cellcolor{gray!40} $0.25\,(.18)$ \\

\hline
\multicolumn{4}{c}{Probability Tutor}\\
\hline
$Pre$ &  $74.8\,(14)$ &  $74.2\,(16)$ & \cellcolor{gray!40} $75.8\,(15)$ \\
$Iso$-$Post$ &  $90.4\,(10)$ &  $65.3\,(16)$  & \cellcolor{gray!40} $90.6\,(8)$\\
$Iso$-$NLG$ &  $0.29\,(.19)$ &  -$0.02\,(.27)$ & \cellcolor{gray!40} $0.26\,(.17)$ \\
$Post$ &  $89.5\,(15)$ &  $62.5\,(18)$ & \cellcolor{gray!40} $88.8\,(7)$\\
$NLG$ &  $0.26\,(.21)$ &  -$0.08\, (.3)$ & \cellcolor{gray!40} $0.24\,(.15)$ \\

\Xhline{4\arrayrulewidth}
\end{tabular} 
\end{center} 
\end{table}
\endgroup

\vskip -0.05in

Table \ref{groupSummary} compares the groups' performance across the two tutors showing the mean and standard deviation of pre- and post-test scores, isomorphic scores, and the learning outcome in terms of the normalized learning gain $(NLG)$ defined as $(NLG = \frac{Post - Pre}{\sqrt{100 - Pre}})$, where $100$ is the maximum test score. We refer to pre-test, post-test and NLG scores as $Pre$, $Post$ and $NLG$, respectively. On both tutors, a one-way ANOVA found no significant difference on $Pre$ between the groups.

To measure the improvement on isomorphic problems, repeated measures ANOVA tests were conducted using \{$Pre$, $Iso$-$Post$\} as factor. Results showed that $Exp$ and $StrTime$ learned significantly with $\mathit{p} <0.0001$ on both tutors, while $Ctrl$ did not perform significantly higher on $Iso$-$Post$ than $Pre$ on both tutors. These findings verify the RFC's accuracy, as $StrTime$ learned significantly on both tutors, while $Ctrl$ did not, despite both receiving no intervention.

A comprehensive comparison between the three groups was essential to evaluate our intervention. On the logic tutor, A one-way ANCOVA using $Pre$ as covariate and group as factor found a significant effect on $Post$: $\mathit{F}(2,117) = 14.5,\, \mathit{p} < .0001,\,\mathit{\eta}^2 = .18$. Subsequent post-hoc analyses with Bonferroni correction $(\alpha=.05/3)$ revealed that $Exp$ and $StrTime$ significantly outperformed $Ctrl$: $\mathit{t}(93) = 3.8,\, \mathit{p} < .001$ and $\mathit{t}(70) = 3.9,\, \mathit{p} < .001$, respectively. Similar patterns were observed on $NLG$ using ANOVA and the post-hoc comparisons.

On the probability tutor, a one-way ANCOVA using $Pre$ as covariate and group as factor showed a significant effect on $Post$: $\mathit{F}(2,117) = 48.1,\, \mathit{p} < .0001,\,\mathit{\eta}^2 = .35$. Follow-up pairwise comparisons with Bonferroni adjustment showed that $Exp$ and $StrTime$ significantly surpassed $Ctrl$: $\mathit{t}(93) = 6.1,\, \mathit{p} < .0001$ and $\mathit{t}(70) = 5.9,\, \mathit{p} < .0001$, respectively. Similar results were found on $NLG$ using ANOVA and the post-hoc comparisons.

\section{Conclusion}

\vskip -0.05in

We showed that mixing BC with FC on the logic tutor improved the experimental students' learning outcomes, as $Exp$ significantly outperformed $Ctrl$ on logic and on a probability tutor that only supports BC. Additionally, $Exp$ caught up with $StrTime$ on both tutors suggesting that $Exp$ students are prepared for future learning \cite{bransford1999transferRethinking} as they acquired BC mastery skills on logic and transferred them to probability, where they received no intervention. There is at least one caveat in our study. The probability tutor supported only one strategy. A more convincing testbed would be having the tutors support both strategies. The future work involves implementing FC on the probability tutor.

\vskip 0.05in
\noindent{\textbf{Acknowledgments:}} This research was supported by the NSF Grants: 1660878, 1651909, 1726550 and 2013502.

\bibliographystyle{splncs04}
\bibliography{academia_refs}

\end{document}